\documentstyle[preprint,aps,floats,epsfig]{revtex}
\tightenlines

\def\Journal#1#2#3#4{{#1} {\bf #2}, #3 (#4)}

\def\IBID{{\em ibid.}}
\def\NP{Nucl. Phys.}
\def\PLB{Phys. Lett. B}
\def\PRC{Phys. Rev. C}
\def\PRD{Phys. Rev. D}
\def\PRL{Phys. Rev. Lett.}
\def\PRT{Phys. Rep.}
\def\ZPA{Z. Phys. A}
\def\ZPC{Z. Phys. C}

\begin{document}
\title{Model for $J/\psi$ absorption in hadronic matter}
\bigskip
\author{Ziwei Lin and C. M. Ko}
\address{Cyclotron Institute and Physics Department, Texas A\&M University,
College Station, Texas 77843-3366}
\maketitle

\begin{abstract}
The cross sections for $J/\psi$ absorption by $\pi$ and $\rho$ mesons 
are studied in a meson-exchange model that includes not only 
pseudoscalar-pseudoscalar-vector-meson couplings but also
three-vector-meson and four-point couplings. 
We find that they are much larger than in a previous study where only 
pseudoscalar-pseudoscalar-vector-meson couplings were considered.
Including form factors at interaction vertices, the $J/\psi$ absorption 
cross sections $\sigma_{\pi \psi}$ and $\sigma_{\rho \psi}$ are found to have 
values on the order of $7$ mb and $3$ mb, respectively. 
Their thermal averages in hadronic matter at temperature $T=150$ MeV are, 
respectively, about $1$ mb and $2$ mb.

\medskip
\noindent PACS number(s): 25.75.-q, 14.40.Gx, 13.75.Lb
\end{abstract}

\section{Introduction}

A dense partonic system, often called the quark-gluon plasma (QGP),
is expected to be formed in heavy ion collisions at the Relativistic 
Heavy Ion Collider (RHIC), which will soon start to operate at the 
Brookhaven National Laboratory. Of all experimental observables that
are sensitive to the presence of the QGP, charmonium is among
the most promising ones. In particular, the dissociation of charmoniums 
in QGP due to color screening would lead to a reduction 
of their production in relativistic heavy ion collisions. The
suppression of charmonium production in these collisions has thus 
been proposed as a possible signature for the formation of QGP 
\cite{matsui}. Extensive experimental and theoretical efforts 
have been devoted to study this phenomenon 
\cite{early,su,na50,quant,review}. However, available experimental 
data on $J/\psi$ suppression in colliding systems ranging from $pA$ 
to S+U are consistent with the scenario that charmoniums are absorbed by 
target and projectile nucleons with a cross section of about $7$ mb 
\cite{quant}.  Only in recent data from the Pb+Pb collision at 
$P_{\rm lab}=158$ GeV$/c$ in the NA50 experiment at CERN \cite{na50}
is there a large additional $J/\psi$ suppression in high $E_{\rm T}$ events, 
which requires the introduction of other absorption mechanisms.
While there are suggestions that this anomalous suppression may be due to 
the formation of QGP \cite{qgp1,qgp2}, other more 
conventional mechanisms based on $J/\psi$ absorption by comoving 
hadrons have also been proposed as a possible explanation
\cite{conv1,conv2}. Since the latter depends on the values of $J/\psi$ 
absorption cross sections by hadrons, which are not known empirically, 
it is important to have better knowledge of the interactions 
between charmonium states and hadrons in order
to understand the nature of the observed anomalous charmonium suppression.

Knowledge of $J/\psi$ absorption cross sections by hadrons is
also useful in estimating the contribution of $J/\psi$ production
from charm mesons in the hadronic matter formed in relativistic
heavy ion collisions. Since the charm meson to $J/\psi$ ratio
in proton-proton collisions increases with energy, it has
been shown that $J/\psi$ production from hadronic matter may
not be negligible in heavy ion collisions at the Large Hadronic Collider
energies \cite{cmk,redlich}. To use $J/\psi$ suppression as a signature
for the formation of QGP in these collisions thus requires 
the understanding of both $J/\psi$ absorption and production
in hadronic matter.

Various approaches have been used in evaluating the charmonium absorption
cross sections by hadrons. In one approach, the quark-exchange model has 
been used. An earlier study based on this model 
by Martins, Blaschke, and Quack \cite{blaschke} has shown
that the $J/\psi$ absorption cross section $\sigma_{\pi \psi}$
by pions has a peak value of about $7$ mb at
$E_{\rm kin}\equiv \sqrt s-m_\pi-m_\psi \simeq 0.8$ GeV, but a 
recent study by Wong, Swanson, and Barnes \cite{wong} gives a peak value 
of only $\sigma_{\pi \psi} \sim 1$ mb at the same $E_{\rm kin}$ region.  
In the perturbative QCD approach,  Kharzeev and Satz \cite{ope} 
have studied the dissociation of charmonium bound states by energetic 
gluons inside hadrons. They have predicted that the dissociation 
cross section increases monotonously with $E_{\rm kin}$ and has a 
value of only about $0.1$ mb around $E_{\rm kin}\sim 0.8$ GeV. 
In the third approach, meson-exchange models based on hadronic 
effective Lagrangians have been used. Using 
pseudoscalar-pseudoscalar-vector-meson couplings (PPV couplings), 
Matinyan and M\"uller \cite{muller} have found 
$\sigma_{\pi \psi} \simeq 0.3$ mb at $E_{\rm kin}=0.8$ GeV.
In a more recent study, Haglin \cite{haglin} has included also the
three-vector-meson couplings (VVV couplings) and four-point couplings 
(or contact terms), 
and obtained much larger values of $J/\psi$ absorption cross 
sections. Large discrepancies in the magnitude of $\sigma_{\pi \psi}$ 
(as well as $\sigma_{\rho \psi}$) thus exist among the predictions from
these three approaches, and further theoretical studies are needed.
In the present study, we use a meson-exchange model as in Ref. 
\cite{haglin} but treat differently the VVV and four-point couplings 
in the effective Lagrangian and also take into account the
effect of form factors at interaction vertices.

Our paper is organized as follows. In Sec. \ref{sec_lagn}, we introduce 
the effective hadronic Lagrangian that we use to obtain the relevant 
interactions among $J/\psi$ and hadrons. The cross sections for 
$J/\psi$ absorption by $\pi$ and $\rho$ mesons are then evaluated. 
The amplitudes for the coherent sum of individual diagrams are checked 
to ensure that the hadronic current is conserved in the limit of zero vector 
meson masses.  
We then show in Sec. \ref{sec_num} the numerical 
results for the cross sections and their dependence on the form factors 
at interaction vertices. 
In Sec. \ref{sec_sum}, we compare our results with other models, and give 
more discussions on form factors and the effect due to the finite
$\rho$ meson width. A summary is also given in this section. 
In Appendix A, we discuss the determination of coupling constants
based on the vector meson dominance model. 
More detailed comparisons with the approach used in Ref.\cite{haglin} 
are given in Appendix B. 

\section{$J/\psi$ absorption in Hadronic Matter}
\label{sec_lagn}

\subsection{Effective Lagrangian}

The free Lagrangian for pseudoscalar and vector mesons 
in the limit of SU(4) invariance can be written as
\begin{eqnarray}
{\cal L}_0= {\rm Tr} \left ( \partial_\mu P^\dagger \partial^\mu P \right )
-\frac{1}{2} {\rm Tr} \left ( F^\dagger_{\mu \nu} F^{\mu \nu} \right )~,
\label{lagn0}
\end{eqnarray}
where $F_{\mu \nu}=\partial_\mu V_\nu-\partial_\nu V_\mu$, 
and $P$ and $V$ denote, respectively, the properly normalized 
$4\times 4$ pseudoscalar and vector meson matrices in SU(4) \cite{ds}:
\begin{eqnarray} P&=&
\frac{1}{\sqrt 2}\left (
\begin{array}{cccc}
\frac{\pi^0}{\sqrt 2}+\frac{\eta}{\sqrt 6} +\frac{\eta_c}{\sqrt {12}}
& \pi^+ & K^+ & \bar {D^0} \\
\pi^- & -\frac{\pi^0}{\sqrt 2}+\frac{\eta}{\sqrt 6}
+\frac{\eta_c}{\sqrt {12}} & K^0 & D^- \\
K^- & \bar {K^0} & -\sqrt {\frac{2}{3}}\eta 
+\frac{\eta_c}{\sqrt {12}} & D_s^- \\
D^0 & D^+ & D_s^+ & -\frac{3\eta_c}{\sqrt {12}} 
\end{array}
\right ) \;, \nonumber \\[2ex]
V&=&\frac{1}{\sqrt 2}\left (
\begin{array}{cccc}
\frac{\rho^0}{\sqrt 2}+\frac{\omega^\prime}{\sqrt 6}
+\frac{J/\psi}{\sqrt {12}} & \rho^+ & K^{*+} & \bar {D^{*0}} \\
\rho^- & -\frac{\rho^0}{\sqrt 2}+\frac{\omega^\prime}{\sqrt 6} 
+\frac{J/\psi}{\sqrt {12}} & K^{*0} & D^{*-} \\
K^{*-} & \bar {K^{*0}} & -\sqrt {\frac{2}{3}}\omega^\prime
+\frac{J/\psi}{\sqrt {12}} & D_s^{*-} \\
D^{*0} & D^{*+} & D_s^{*+} & -\frac{3J/\psi}{\sqrt {12}}
\end{array}
\right ) \;.  
\label{pv}
\end{eqnarray}
To obtain the couplings between pseudoscalar mesons and vector mesons, 
we introduce the minimal substitution
\begin{eqnarray}
\partial_\mu P &\rightarrow& {\cal D}_\mu P= \partial_\mu P
-\frac{ig}{2} \left [V_\mu P \right ]~, \label{ms1} \\
F_{\mu \nu} &\rightarrow& 
\partial_\mu V_\nu-\partial_\nu V_\mu -\frac{ig}{2} 
\left [ V_\mu, V_\nu \right ]~.
\label{ms2}
\end{eqnarray}
The effective Lagrangian is then given by 
\begin{eqnarray}
{\cal L}&=& {\cal L}_0 +\frac{ig}{2} {\rm Tr} 
\left ( \partial^\mu P \left [P^\dagger, V^\dagger_\mu \right ] 
+\partial^\mu P^\dagger \left [P, V_\mu \right ] \right ) 
-\frac{g^2}{4} {\rm Tr} \left ( \left [ P^\dagger, V^\dagger_\mu \right ]
\left [ P, V^\mu \right ] \right ) \nonumber \\
&+&\frac{ig}{2} {\rm Tr} 
\left ( \partial^\mu V^\nu \left [V^\dagger_\mu, V^\dagger_\nu \right ] 
+\partial_\mu V^\dagger_\nu \left [V^\mu, V^\nu \right ] \right ) 
+\frac{g^2}{8} {\rm Tr} 
\left ( \left [V^\mu, V^\nu \right ] 
\left [V^\dagger_\mu, V^\dagger_\nu \right ] \right )~.
\label{lagn1}
\end{eqnarray}
The hermiticity of $P$ and $V$ reduces this to 
\begin{eqnarray}
{\cal L}&=& {\cal L}_0 + ig {\rm Tr} 
\left ( \partial^\mu P \left [P, V_\mu \right ] \right ) 
-\frac{g^2}{4} {\rm Tr} 
\left ( \left [ P, V_\mu \right ]^2 \right ) \nonumber \\
&+& ig {\rm Tr} \left ( \partial^\mu V^\nu \left [V_\mu, V_\nu \right ] 
\right ) 
+\frac{g^2}{8} {\rm Tr} \left ( \left [V_\mu, V_\nu \right ]^2 \right )~.
\label{lagn2}
\end{eqnarray}
Since the SU(4) symmetry is explicitly broken by hadron masses, terms
involving hadron masses are added to Eq.(\ref{lagn2}) using the
experimentally determined values. 

\subsection{Effective Lagrangians relevant for $J/\psi$ absorption}

Expanding the Lagrangian in Eq.(\ref{lagn2}) explicitly in terms of
the pseudoscalar meson and vector meson matrices shown in Eq.(\ref{pv}),  
we obtain the following interaction 
Lagrangians that are relevant 
for the study of $J/\psi$ absorption by $\pi$ and $\rho$ mesons:
\begin{eqnarray}
{\cal L}_{\pi DD^*}&=&ig_{\pi DD^*}~ 
D^{* \mu} \vec \tau \cdot \left ( \bar D \partial_\mu \vec \pi - 
\partial_\mu \bar D \vec \pi \right ) + {\rm H.c.}~ , \label{pdds}\\ 
{\cal L}_{\psi DD}&=&ig_{\psi DD}~ \psi^\mu 
\left ( D \partial_\mu \bar D -\partial_\mu D \bar D \right ) ~ ,\label{jdd}\\ 
{\cal L}_{\psi D^*D^*}&=& ig_{\psi D^*D^*}~
\left [ \psi^\mu \left ( \partial_\mu D^{* \nu} \bar {D^*_\nu} 
- D^{* \nu} \partial_\mu \bar {D^*_\nu} \right )
+\left ( \partial_\mu \psi^\nu D^*_\nu -\psi^\nu \partial_\mu D^*_\nu \right )
\bar {D^{* \mu}} \right .\nonumber \\ 
&+&\left . D^{* \mu} \left ( \psi^\nu \partial_\mu \bar {D^*_\nu} -
\partial_\mu \psi^\nu \bar {D^*_\nu} \right ) \right ] ~ , \label{jdsds} \\ 
{\cal L}_{\pi \psi DD^*}&=&-g_{\pi \psi DD^*}~
\psi^\mu \left ( D^*_\mu \vec \tau \bar D
+ D \vec \tau \bar {D^*_\mu} \right ) \cdot \vec \pi ~ , \label{pjdds} \\ 
{\cal L}_{\rho DD}&=&ig_{\rho DD}~ \left ( D \vec \tau \partial_\mu \bar D
-\partial_\mu D \vec \tau \bar D \right ) \cdot \vec \rho^\mu ~ ,\nonumber \\ 
{\cal L}_{\rho \psi DD}&=&g_{\rho \psi DD}~ 
\psi^\mu D \vec \tau \bar D \cdot \vec {\rho_\mu} ~ ,\nonumber \\ 
{\cal L}_{\rho D^*D^*}&=&ig_{\rho D^*D^*}~ \left [ 
\left ( \partial_\mu D^{* \nu} \vec \tau \bar {D^*_\nu}
-D^{* \nu} \vec \tau \partial_\mu \bar {D^*_\nu} \right ) \cdot \vec \rho^\mu
+\left ( D^{* \nu} \vec \tau \cdot \partial_\mu \vec \rho_\nu 
-\partial_\mu D^{* \nu} \vec \tau \cdot \vec \rho_\nu \right ) 
\bar {D^{* \mu}} \right .\nonumber \\ 
&+& \left . D^{* \mu} \left ( 
\vec \tau \cdot \vec \rho^\nu \partial_\mu \bar {D^*_\nu}
-\vec \tau \cdot \partial_\mu \vec \rho^\nu \bar {D^*_\nu} \right ) \right ] 
~ ,\nonumber \\ 
{\cal L}_{\rho \psi D^*D^*}&=&g_{\rho \psi D^*D^*}~ \left ( 
\psi^\nu D^*_\nu \vec \tau \bar {D^*_\mu} 
+\psi^\nu D^*_\mu \vec \tau \bar {D^*_\nu} 
-2 \psi_\mu D^{* \nu}\vec \tau \bar {D^*_\nu} \right ) \cdot \vec \rho^\mu ~ .
\end{eqnarray}
In the above, $\vec \tau$ are the Pauli matrices, 
and $\vec \pi$ and $\vec \rho$ denote the pion and rho meson isospin triplets, 
respectively, while $D\equiv (D^0,D^+)$ and $D^*\equiv (D^{*0},D^{*+})$ 
denote the pseudoscalar and vector charm meson doublets, respectively.
We note that exact SU(4) symmetry would give the following relations 
among the coupling constants in the Lagrangian:
\begin{eqnarray}
&&g_{\pi DD^*}=g_{\rho DD}=g_{\rho D^* D^*}=\frac{g}{4}~, ~
g_{\psi DD}=g_{\psi D^* D^*}=\frac{g}{\sqrt 6}~, \nonumber \\
&&g_{\pi \psi DD^*}=g_{\rho \psi D^* D^*}=\frac{g^2}{4 \sqrt 6}~, ~
g_{\rho \psi D D}=\frac{g^2}{2 \sqrt 6}~.  
\label{su4}
\end{eqnarray}

\subsection{$J/\psi$ absorption cross sections}

\begin{figure}
\centerline{\epsfig{file=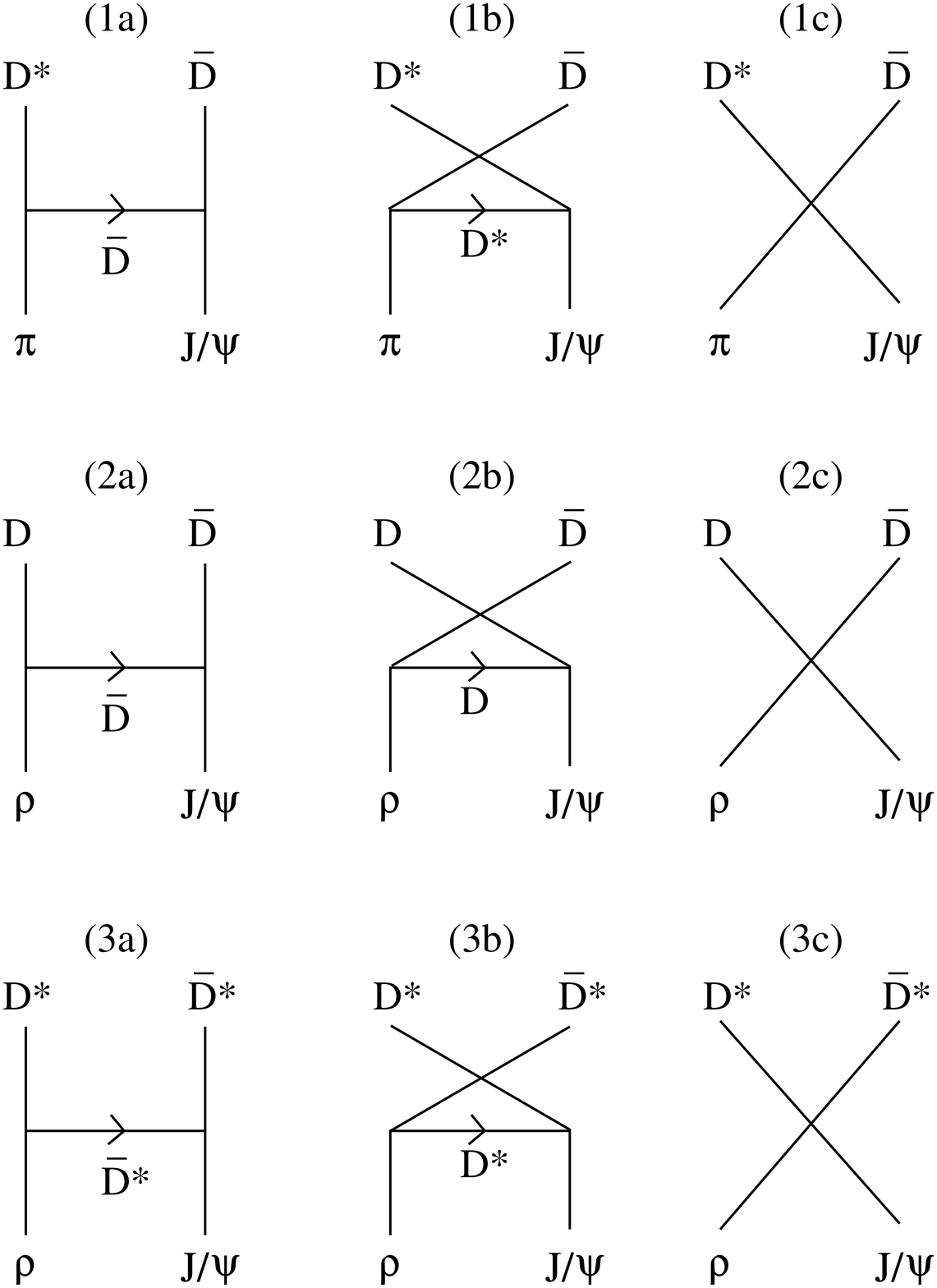,height=3in,width=3in,angle=0}}
\vspace{1cm}
\caption{Diagrams for $J/\psi$ absorption processes:   
1) $\pi \psi \rightarrow D^* \bar D$;
2) $\rho \psi \rightarrow D \bar D$; 
and 3) $\rho \psi \rightarrow D^* \bar {D^*}$. Diagrams for
the process $\pi \psi \rightarrow D \bar {D^*}$ are similar to 
(1a)-(1c) but with each particle replaced by its antiparticle.
}
\label{diagrams}
\end{figure}

The above effective Lagrangians allow us to study the following processes 
for $J/\psi$ absorption by $\pi$ and $\rho$ mesons:
\begin{eqnarray}
\pi \psi \rightarrow D^* \bar D, ~ \pi \psi \rightarrow D \bar {D^*},~
\rho \psi \rightarrow D \bar D, ~ \rho \psi \rightarrow D^* \bar {D^*}. 
\end{eqnarray}
The corresponding diagrams for these processes, except 
the process $\pi \psi \rightarrow D \bar {D^*}$, 
which has the same cross section as the process
$\pi \psi \rightarrow D^* \bar D$, are shown in Fig. \ref{diagrams}. 

The full amplitude for the first process $\pi \psi \rightarrow D^* \bar D$, 
without isospin factors and before summing and averaging over external spins, 
is given by
\begin{eqnarray}
{\cal M}_1 \equiv {\cal M}_1^{\nu \lambda} 
~\epsilon_{2 \nu} \epsilon_{3 \lambda} 
=\left ( \sum_{i=a,b,c} {\cal M}_{1 i}^{\nu \lambda} \right )
\epsilon_{2 \nu} \epsilon_{3 \lambda}, 
\label{m_i}
\end{eqnarray}
with
\begin{eqnarray}
{\cal M}_{1a}^{\nu \lambda}&=& -g_{\pi D D^*} g_{\psi D D}~
(-2p_1+p_3)^\lambda \left (\frac{1}{t-m_D^2} \right ) 
(p_1-p_3+p_4)^\nu, \nonumber \\
{\cal M}_{1b}^{\nu \lambda}&=& g_{\pi D D^*} g_{\psi D^* D^*}~
(-p_1-p_4)^\alpha \left ( \frac{1}{u-m_{D^*}^2} \right )
\left [ g_{\alpha \beta}-\frac{(p_1-p_4)_\alpha (p_1-p_4)_\beta}{m_{D^*}^2}
\right ] \nonumber \\
&\times& \left [ (-p_2-p_3)^\beta g^{\nu \lambda}
+(-p_1+p_2+p_4)^\lambda g^{\beta \nu}
+(p_1+p_3-p_4)^\nu g^{\beta \lambda} \right ] , \nonumber \\
{\cal M}_{1c}^{\nu \lambda}&=& -g_{\pi \psi D D^*}~ g^{\nu \lambda}.
\label{pij}
\end{eqnarray}

Similarly, the full amplitude for the second process 
$\rho \psi \rightarrow D \bar D$ is given by
\begin{eqnarray}
{\cal M}_2 \equiv {\cal M}_2^{\mu \nu} \epsilon_{1 \mu} \epsilon_{2 \nu} 
=\left ( \sum_{i=a,b,c} {\cal M}_{2 i}^{\mu \nu} \right )
~\epsilon_{1 \mu} \epsilon_{2 \nu}, 
\end{eqnarray}
with 
\begin{eqnarray}
{\cal M}_{2a}^{\mu \nu}&=& -g_{\rho D D} g_{\psi D D}~
(p_1-2 p_3)^\mu \left ( \frac{1}{t-m_D^2} \right ) 
(p_1-p_3+p_4)^\nu, \nonumber \\
{\cal M}_{2b}^{\mu \nu}&=& -g_{\rho D D} g_{\psi D D}~
(-p_1+2 p_4)^\mu\left ( \frac{1}{u-m_D^2}\right ) (-p_1-p_3+p_4)^\nu, 
\nonumber \\
{\cal M}_{2c}^{\mu \nu}&=& g_{\rho \psi D D}~ g^{\mu \nu}.
\end{eqnarray}

For the third process $\rho \psi \rightarrow D^* \bar {D^*}$, 
the full amplitude is given by
\begin{eqnarray}
{\cal M}_3
\equiv {\cal M}_3^{\mu \nu \lambda \omega} 
~\epsilon_{1 \mu} \epsilon_{2 \nu} \epsilon_{3 \lambda} \epsilon_{4 \omega} 
=\left ( \sum_{i=a,b,c} {\cal M}_{3 i}^{\mu \nu \lambda \omega} \right )
\epsilon_{1 \mu} \epsilon_{2 \nu} \epsilon_{3 \lambda} \epsilon_{4 \omega}, 
\end{eqnarray}
with
\begin{eqnarray}
{\cal M}_{3a}^{\mu \nu \lambda \omega}&=& g_{\rho D^* D^*} g_{\psi D^* D^*}~
\left [ (-p_1-p_3)^\alpha g^{\mu \lambda} +2 ~p_1^\lambda g^{\alpha \mu}
+2 p_3^\mu g^{\alpha \lambda} \right ] \left ( \frac{1}{t-m_{D^*}^2} \right )
\nonumber \\
&\times& \left [ g_{\alpha \beta}-\frac{(p_1-p_3)_\alpha (p_1-p_3)_\beta}
{m_{D^*}^2} \right ] \left [ -2 p_2^\omega g^{\beta \nu}
+(p_2+p_4)^\beta g^{\nu \omega}
-2 p_4^\nu g^{\beta \omega} \right ] , \nonumber \\
{\cal M}_{3b}^{\mu \nu \lambda \omega}&=& g_{\rho D^* D^*} g_{\psi D^* D^*}~
\left [ -2 p_1^\omega g^{\alpha \mu} + (p_1+p_4)^\alpha g^{\mu \omega}
-2 p_4^\mu g^{\alpha \omega} \right ] \left ( \frac{1}{u-m_{D^*}^2} \right )
\nonumber \\
&\times& \left [ g_{\alpha \beta}-\frac{(p_1-p_4)_\alpha (p_1-p_4)_\beta}
{m_{D^*}^2} \right ] \left [ (-p_2-p_3)^\beta g^{\nu \lambda}
+2 p_2^\lambda g^{\beta \nu}+2 p_3^\nu g^{\beta \lambda}\right ], \nonumber \\
{\cal M}_{3c}^{\mu \nu \lambda \omega}&=& g_{\rho \psi D^* D^*}~
\left ( g^{\mu \lambda} g^{\nu \omega} +g^{\mu \omega} g^{\nu \lambda} 
-2 g^{\mu \nu} g^{\lambda \omega} \right ).
\end{eqnarray}
In the above, $p_j$ denotes the momentum of particle $j$.
We choose the convention that particles $1$ and $2$ represent 
initial-state mesons while particles $3$ and $4$ 
represent final-state mesons on the left and right sides of the diagrams 
shown in Fig. \ref{diagrams}, respectively.
The indices $\mu, \nu, \lambda$, and $\omega$ 
denote the polarization components of external particles while
the indices $\alpha$ and $\beta$ denote those of the exchanged
mesons. 

After averaging (summing) over initial (final) spins 
and including isospin factors, the cross sections 
for the three processes are given by 
\begin{eqnarray}\label{jpion}
\frac {d\sigma_1}{dt}&=& \frac {1}{96 \pi s p_{i,\rm cm}^2} 
{\cal M}_1^{\nu \lambda} {\cal M}_1^{*\nu^\prime \lambda^\prime}
\left ( g_{\nu \nu^\prime}-\frac{p_{2 \nu} p_{2 \nu^\prime}} {m_2^2} \right )
\left ( g_{\lambda \lambda^\prime}
-\frac{p_{3 \lambda} p_{3 \lambda^\prime}} {m_3^2} \right ),
\end{eqnarray}
\begin{eqnarray}
\frac {d\sigma_2}{dt}&=&\frac {1}{288 \pi s p_{i,\rm cm}^2}
{\cal M}_2^{\mu \nu} {\cal M}_2^{*\mu^\prime \nu^\prime}
\left ( g_{\mu \mu^\prime}-\frac{p_{1 \mu} p_{1 \mu^\prime}} {m_1^2} \right )
\left ( g_{\nu \nu^\prime}-\frac{p_{2 \nu} p_{2 \nu^\prime}} {m_2^2} \right ),
\end{eqnarray}
\begin{eqnarray}
\frac {d\sigma_3}{dt}&=&\frac {1}{288 \pi s p_{i,\rm cm}^2} 
{\cal M}_3^{\mu \nu \lambda \omega} 
{\cal M}_3^{*\mu^\prime \nu^\prime \lambda^\prime \omega^\prime}
\left ( g_{\mu \mu^\prime}-\frac{p_{1 \mu} p_{1 \mu^\prime}} {m_1^2} \right )
\left ( g_{\nu \nu^\prime}-\frac{p_{2 \nu} p_{2 \nu^\prime}} {m_2^2} \right ) 
\nonumber \\
&\times& \left ( g_{\lambda \lambda^\prime}
-\frac{p_{3 \lambda} p_{3 \lambda^\prime}} {m_3^2} \right )
\left ( g_{\omega \omega^\prime}
-\frac{p_{4 \omega} p_{4 \omega^\prime}} {m_4^2} \right ) ,
\label{m3}
\end{eqnarray}
with $s=(p_1+p_2)^2$, and  
\begin{eqnarray}
p_{i,\rm cm}^2=\frac {\left [ s-(m_1+m_2)^2 \right ]
\left [ s-(m_1-m_2)^2 \right ]}{4s}
\end{eqnarray}
is the squared momentum of initial-state mesons in the 
center-of-momentum (c.m.) frame. 

\subsection{Current conservation}
\label{sec_cc} 

The effective Lagrangian in Eq. (\ref{lagn2}) is
generated by minimal substitution, which is equivalent
to treating vector mesons as gauge particles.  To preserve the gauge
invariance in the limit of zero vector meson masses thus leads to 
both VVV and four-point couplings in the Lagrangian.
The gauge invariance also results in current conservation; i.e., 
in the limit of zero vector meson masses, degenerate pseudoscalar
meson masses, and SU(4) invariant coupling constants, one has
\begin{eqnarray}\label{current} 
{\cal M}_n^{\lambda_k \dots \lambda_l}~ p_{j \lambda_j}=0~, 
\end{eqnarray}
where the index $\lambda_j$ denotes the external vector meson $j$
in process $n$ shown in Fig. \ref{diagrams}.
This then requires, e.g., 
${\cal M}_1^{\nu \lambda} p_{3 \lambda}=0$ 
and ${\cal M}_3^{\mu \nu \lambda \omega} p_{2 \nu}=0$. 
In Appendix B, we shall explicitly check that the amplitudes given 
in Eq. (\ref{pij}),
as an example, satisfy the requirement of current conservation.  

\section{Numerical Results}
\label{sec_num}

The coupling constant $g_{\pi DD^*}$ can be determined from the $D^*$
decay width \cite{dwidth}, and this gives $g_{\pi DD^*}=4.4$.
Using the vector meson dominance (VMD) model, we can 
determine other three-point coupling constants. As shown in 
Appendix A, their values are
\begin{eqnarray}
g_{\rho DD}=g_{\rho D^* D^*}=2.52~,~
g_{\psi DD}=g_{\psi D^* D^*}=7.64~.  
\end{eqnarray}
For the four-point coupling constants, there is no empirical information,
and we thus use the SU(4) relations to determine their values in terms of the
three-point coupling constants, i.e., 
\begin{eqnarray}
g_{\pi \psi DD^*}= g_{\pi DD^*} g_{\psi DD}, ~
g_{\rho \psi D D}= 2~g_{\rho DD} g_{\psi DD}, ~
g_{\rho \psi D^* D^*}=g_{\rho D^* D^*} g_{\psi D^* D^*}.  
\end{eqnarray}
To obtain analytical expressions for the cross sections so that 
they can be directly included into a computer code for numerical 
calculations, we have used the software package FORM \cite{form} 
to contract all Lorentz indices in Eq. (\ref{m3}). 

\subsection{Without form factors}

\begin{figure}
\centerline{\epsfig{file=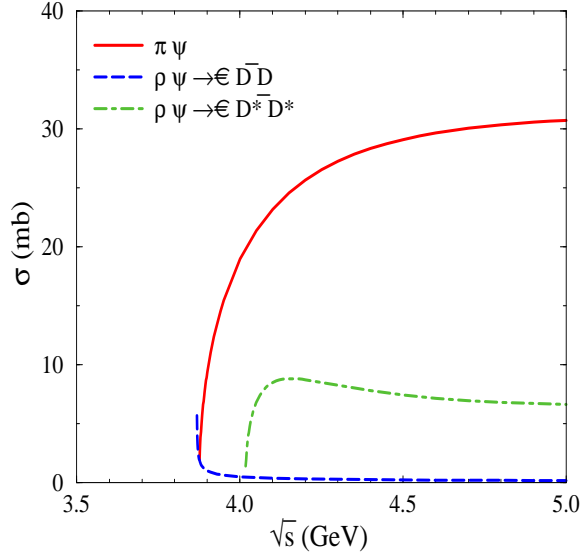,height=3in,width=3in,angle=0}}
\caption{
$J/\psi$ absorption cross section (without form factors) 
as a function of the c.m. energy of initial-state mesons.
The solid curve represents the total contribution from both 
$\pi \psi \rightarrow D^* \bar D$
and $\pi \psi \rightarrow \bar {D^*} D$ processes.
}
\label{sigma}
\end{figure}

Figure \ref{sigma} shows the cross section of $J/\psi$ absorption by 
$\pi$ and $\rho$ mesons as a function of 
the c.m. energy $\sqrt s$ of the two initial-state mesons. 
The cross section $\sigma_{\pi\psi}$, shown by the solid curve, 
includes contributions from both $\pi \psi \rightarrow D \bar {D^*}$
and $\pi\psi\to D^*{\bar D}$, which have same cross sections.
It is seen that the three $J/\psi$
absorption cross sections have very different energy dependence near 
the threshold energy, ${\rm max}(m_1+m_2,m_3+m_4)$.  While 
$\sigma_{\pi \psi}$ increases monotonously with c.m. energy,  
the cross section for the process $\rho\psi\to D{\bar D}$
decreases rapidly with c.m. energy, and that for the process 
$\rho\psi\to D^*{\bar D^*}$ changes little with c.m. energy after an initial
rapid increase near the threshold.

The thermal average of these cross sections 
in a hadronic matter at temperature $T$ is given by 
\begin{eqnarray} 
\langle \sigma v \rangle 
&=&\frac{\int_{z_0}^{\infty} dz \left [z^2-(\alpha_1+\alpha_2)^2
\right ] \left [z^2-(\alpha_1-\alpha_2)^2 \right ] K_1(z) ~
\sigma (s=z^2{\rm T}^2)}
{4 \alpha_1^2 K_2(\alpha_1)\alpha_2^2 K_2(\alpha_2)} ~,  
\end{eqnarray} 
where $\alpha_i=m_i/{\rm T}$ ($i=1$ to $4$), 
$z_0={\rm max}(\alpha_1+\alpha_2,\alpha_3+\alpha_4)$, 
$K_n$'s are modified Bessel functions, and 
$v$ is the relative velocity of initial-state 
particles in their collinear frame, 
i.e.,
\begin{eqnarray} 
v=\frac {\sqrt {(k_1\cdot k_2)^2-m_1^2 m_2^2}}{E_1 E_2}. 
\end{eqnarray}

\begin{figure}
\centerline{\epsfig{file=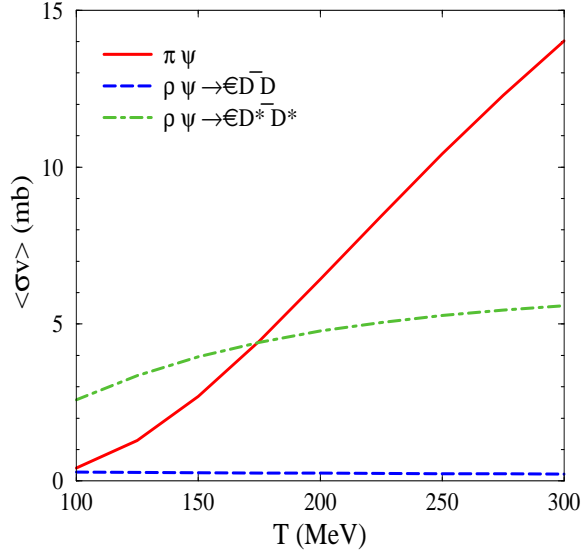,height=3in,width=3in,angle=0}}
\caption{Thermal average of $J/\psi$ absorption cross section 
(without form factors) as a function of temperature T. 
}
\label{sv}
\end{figure}

As shown in Fig. \ref{sv},
$\langle \sigma_{\pi \psi}v \rangle$ increases 
with increasing temperature, but $\langle \sigma_{\rho \psi}v \rangle$ 
varies only moderately with temperature. The contribution of
the process $\rho\psi\to D{\bar D}$ to 
$\langle \sigma_{\rho \psi}v \rangle$ is seen to decrease slightly
with temperature. 
These features can be understood from the energy dependence of 
the cross sections shown in Fig. \ref{sigma} and the difference in their
kinematic thresholds (i.e., $m_3+m_4-m_1-m_2$),
which are about $0.64, -0.14$, and $0.15$ GeV for 
the processes $\pi\psi\to D^*{\bar D}(D{\bar D^*})$,
$\rho\psi\to D{\bar D}$, and $\rho\psi\to D^*{\bar D^*}$, 
respectively.  The process $\pi\psi\to D^*{\bar D}(D{\bar D^*})$
has the highest threshold, while the process $\rho\psi\to D{\bar D}$ is 
exothermic and thus has no threshold. With a pion in the initial state, 
the process $\pi\psi\to D^*{\bar D}(D{\bar D^*})$
requires very energetic pions to overcome the high energy
threshold and thus has a small thermal average 
$\langle \sigma_{\pi \psi}v \rangle$ 
at low temperature. At higher temperature not only are there 
more energetic pions that are able to overcome 
the kinematic threshold but also the cross section for the
process $\pi\psi\to D^*{\bar D}(D{\bar D^*})$ 
increases with the c.m. energy as shown in Fig. \ref{sigma};  
$\langle \sigma_{\pi \psi}v \rangle$ thus
increases strongly with temperature.
For the process $\rho\psi\to D{\bar D}$, on the other hand, 
its contribution to the thermal average 
$\langle \sigma_{\rho \psi}v \rangle$ decreases with temperature
because with increasing temperature there are fewer rho mesons 
at low energy, which contribute the largest cross section.
The contribution of the process $\rho\psi\to D^*{\bar D^*}$ to
$\langle \sigma_{\rho \psi}v \rangle$ 
changes slowly with temperature as a result of both the small threshold 
and the fact that the cross section only weakly depends on the c.m. energy.

Compared with the results of Matinyan and M\"uller \cite{muller}, 
we see that the inclusion of the VVV and four-point couplings increases 
$\sigma_{\pi \psi}$ by an order of magnitude. For the process
$\rho \psi \rightarrow D \bar D$,
the decrease of its cross section after including 
four-point couplings is due to their destructive interference 
with the PPV coupling terms.
The process $\rho \psi \rightarrow D^* \bar {D^*}$ is entirely due to 
VVV and four-point couplings and is seen to have 
a much larger cross section than that for the process 
$\rho \psi \rightarrow D \bar D$.
As a result, our effective Lagrangian including 
the VVV and four-point couplings also significantly increases 
$\sigma_{\rho \psi}$.

\subsection{With form factors}
\label{sec_ff}

To take into account the composite nature of hadrons, 
form factors need to be introduced at interaction vertices. 
Unfortunately, there is no empirical information on form factors 
involving charmoniums and charm mesons. 
We thus take the form factors as the usual monopole form 
at the three-point $t$ channel and $u$ channel vertices, i.e., 
\begin{eqnarray}
f_3=\frac {\Lambda^2}{\Lambda^2+{\bf q}^2},
\end{eqnarray}
where $\Lambda$ is a cutoff parameter, and 
${\bf q}^2$ is the squared three momentum transfer in the c.m. frame, 
given by $({\bf p_1}-{\bf p_3})_{\rm c.m.}^2$ 
and $({\bf p_1}-{\bf p_4})_{\rm c.m.}^2$ 
for $t$ and $u$ channel processes, respectively. 
We assume that the form factor at four-point vertices has the form
\begin{eqnarray}
f_4=\left ( \frac {\Lambda_1^2}{\Lambda_1^2+\bar {{\bf q}^2} } \right )
\left ( \frac {\Lambda_2^2}{\Lambda_2^2+\bar {{\bf q}^2} } \right ),
\end{eqnarray}
where $\Lambda_1$ and $\Lambda_2$ are
the two different cutoff parameters at the three-point vertices 
present in the process with the same initial and final particles, 
and $\bar {{\bf q}^2}$ is the average value 
of the squared three momentum transfers in $t$ and $u$ channels:   
\begin{eqnarray}
\bar {{\bf q}^2} = \frac {\left [ ({\bf p}_1-{\bf p}_3)^2
+({\bf p}_1-{\bf p}_4)^2 \right ]_{\rm c.m.}}{2}=p_{i,\rm c.m.}^2
+p_{f,\rm c.m.}^2~.
\end{eqnarray}
For simplicity, 
we use the same value for all cutoff parameters, i.e.,
\begin{eqnarray}
\Lambda_{\pi D D^*}=\Lambda_{\rho DD}=\Lambda_{\rho D^* D^*}
=\Lambda_{\psi DD}=\Lambda_{\psi D^* D^*}\equiv \Lambda, 
\end{eqnarray}
and choose $\Lambda$ as either $1$ or $2$ GeV 
to study the uncertainties due to form factors.

\begin{figure}
\centerline{\epsfig{file=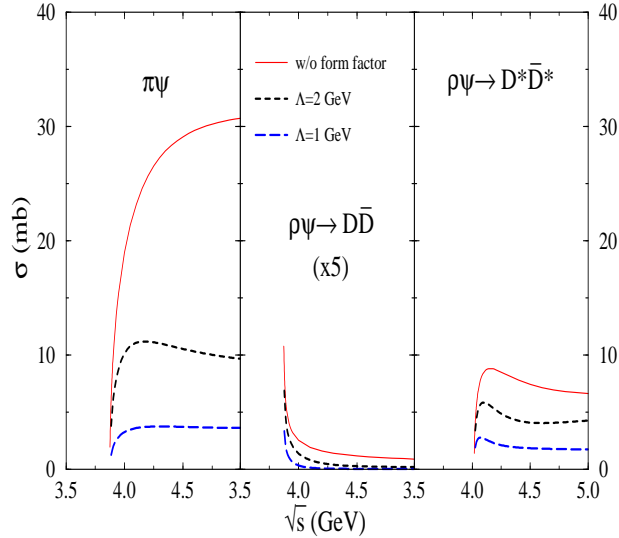,height=3in,width=3in,angle=0}}
\caption{
$J/\psi$ absorption cross section
as a function of the c.m. energy of initial-state mesons
with and without form factors. 
}
\label{sigma-ff}
\end{figure}

\begin{figure}
\centerline{\epsfig{file=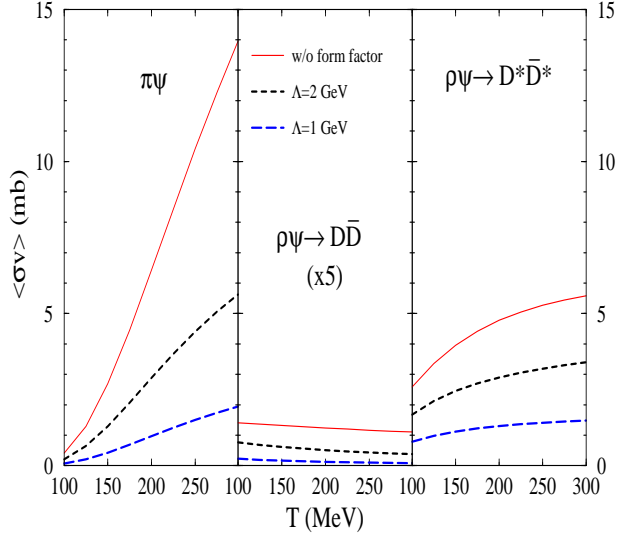,height=3in,width=3in,angle=0}}
\caption{
Thermal-averaged cross section of $J/\psi$ absorption 
as a function of temperature T with and without form factors. 
}
\label{sv-ff}
\end{figure}

Figure \ref{sigma-ff} shows the cross section as a function of the c.m. energy
without and with form factors.  It is seen that form factors strongly 
suppress the cross sections and thus cause large uncertainties in their
values. However, the $J/\psi$ absorption 
cross sections remain appreciable after including form factors 
at interaction vertices. 
The values for $\sigma_{\pi \psi}$ and $\sigma_{\rho \psi}$ 
are roughly $7$ mb and $3$ mb, respectively, 
and are comparable to those used in phenomenological
studies of $J/\psi$ absorption by comoving hadrons in relativistic
heavy ion collisions \cite{conv1,conv2,comover}.
The thermal average of $J/\psi$ absorption cross sections 
with and without form factors is shown in Fig. \ref{sv-ff}. 
At the temperature of $150$ MeV, for example, 
$\langle \sigma_{\pi \psi}v \rangle$ and $\langle \sigma_{\rho \psi}v \rangle$ 
are about $1$ mb and $2$ mb, respectively. 

\section{discussions and summary}
\label{sec_sum}

In our study, the effective Lagrangian shown in Eq. (\ref{lagn2}) 
is obtained from applying the minimal substitution 
of Eq. (\ref{ms1}) and Eq. (\ref{ms2}) to the free Lagrangian.  The resulting
PPV, VVV, and four-point (PPVV and VVVV) interaction Lagrangians
in Eq. (\ref{lagn2}) are exactly the same as those in the chiral 
Lagrangian approach \cite{song}
\footnote{Before checking the identity of our effective 
Lagrangian to the corresponding ones
in the chiral Lagrangian approach of Ref.\cite{song},
one should take notice of the different normalizations for 
the coupling constant $g$ and the meson matrices
as well as the following typos in that paper. 
The first part of Eq. (A2) in Ref.\cite{song} should be  
$U={\rm exp} \left [ i\frac{\sqrt 2}{f_\pi} \phi \right ]$, 
Eq. (A3) should be  
$D_\mu U= \partial_\mu U -ig A^L_\mu U + ig U A^R_\mu$, 
and the first part of Eq. (A5) should be 
${\cal L}^{(3)}_{V\phi \phi}=-ig/2~{\rm Tr} \partial_\mu \phi 
\left [V^\mu, \phi \right ] + \cdots$.}.  
They are, however, different from those used by Haglin \cite{haglin}. 
The differences are shown in detail in Appendix B. 

Values of the $J/\psi$ absorption cross sections by hadrons
obtained in our model are comparable to 
those from Martins, Blaschke, and Quack\cite{blaschke}, 
Haglin \cite{haglin}, and Wong, Swanson, and Barnes \cite{wong}, 
but are much larger than those from Kharzeev and Satz \cite{ope}  
and Matinyan and M\"uller \cite{muller}. 
As shown in Fig. \ref{sigma} and Fig. \ref{sv},
our results without form factors are much larger than those from 
Matinyan and  M\"uller \cite{muller} because the latter only included
pseudoscalar-pseudoscalar-vector-meson couplings. 
As to the energy dependence of $J/\psi$ absorption
cross sections, our results shown in Fig. \ref{sigma} 
for the case without form factors 
are similar to those from Matinyan and M\"uller \cite{muller} and 
Haglin \cite{haglin}, which are also based on effective hadronic Lagrangians.
Including form factors weakens the energy dependence of the absorption
cross sections as shown in Fig. \ref{sigma-ff}. However, the decrease of
the absorption cross sections with energy is still not as fast as
in quark-exchange models \cite{blaschke,wong}.  This difference 
could be due to the fact that meson interactions in our 
effective hadronic Lagrangian approach involve derivative couplings, 
leading thus to a strong momentum dependence in the matrix elements,
while the nonrelativistic potential used in the quark-exchange model 
does not have an explicit momentum dependence. 
Including the relativistic corrections to the quark-quark potential will be 
useful for further studying the energy dependence of the $J/\psi$ absorption 
cross sections in the quark-exchange model.

Form factors involving charm mesons 
introduce significant uncertainties to our model based on 
hadronic effective Lagrangians, because there is 
little experimental information available.
Four-point vertices appear in all processes in our study.
If all vector mesons are massless, 
it is possible to determine the form factor at a four-point 
vertex once form factors at three-point vertices are chosen \cite{ff}.
This is achieved through gauge invariance by 
requiring current conservation for the total amplitude 
that includes the form factors. Since the uncertainty 
of form factors involving charm mesons is already large 
for three-point vertices and the gauge invariance is only exact 
when all vector mesons are massless, we choose not to follow this
more involved approach. Instead, we show the uncertainties  
due to form factors by using two different values for 
the cutoff parameters.

We have used the centroid value for the $\rho$ meson mass in this study. 
Since the $\rho$ meson width in vacuum is large ($151$ MeV), 
the threshold behavior of $\rho \psi$ processes may change significantly  
with the $\rho$ meson mass. 
E.g., a rho meson with mass below $630$ MeV changes the
process $\rho\psi\to D{\bar D}$ from exothermic to endothermic,
and the energy dependence of the cross section near the threshold
may thus change from fast decreasing (the dashed curve) shown 
in Fig. \ref{sigma} to fast rising (similar to the dot-dashed curve).
On the other hand, a rho meson with mass above $920$ MeV changes 
the process $\rho\psi\to D^*{\bar D^*}$ 
from endothermic to exothermic. We thus expect that the final value
of the $J/\psi$ absorption cross section by rho mesons will be different
once the $\rho$ meson width is considered. However, 
the $\rho$ meson spectral function is further modified 
in the hadronic matter produced in heavy ion collisions \cite{rhoth,rhoexp},
so the effects of the rho meson width on $J/\psi$ absorption in hadronic
matter are more involved. 
We therefore leave the effect of the $\rho$ meson width on charmonium 
absorption to a future study.

Finally, vector mesons are treated as gauge particles in our approach.
Since the SU(4) symmetry is not exact, it is not clear to what extent 
they can be treated as gauge particles. 
An alternative approach \cite{chan} based on both chiral symmetry and 
heavy quark effective theory may be useful in understanding the 
meson-exchange model we have used here. 

In summary, we have studied the $J/\psi$ absorption cross sections 
by $\pi$ and $\rho$ mesons in a meson-exchange model that
includes pseudoscalar-pseudoscalar-vector-meson couplings,
three-vector-meson couplings, and four-point couplings.
We find that these cross sections 
have much larger values than in a previous study, 
where only pseudoscalar-pseudoscalar-vector-meson couplings were considered.
Including form factors at the interaction vertices, 
the values for $\sigma_{\pi \psi}$ and $\sigma_{\rho \psi}$ 
are on the order of $7$ mb and $3$ mb, respectively,  
and their thermal averages at the temperature of $150$ MeV 
are roughly $1$ mb and $2$ mb, respectively.   
These values are comparable to those used in the phenomenological studies
of $J/\psi$ absorption in relativistic heavy ion collisions. 
Our results thus suggest that the absorption of $J/\psi$ by comoving hadrons 
may play an important role in the observed suppression.

\section*{Acknowledgments} 

We thank K. Haglin and C. Y. Wong for helpful communications.
This work was supported in part by the National Science Foundation under 
Grant No. PHY-9870038, the Welch Foundation under Grant No. A-1358,
and the Texas Advanced Research Program under Grants Nos. FY97-010366-0068
and FY99-010366-0081.

\pagebreak
\section*{Appendix A}
\setcounter{equation}{0}
\def\theequation{A\arabic{equation}}

In this appendix, 
we determine the values of the coupling constants within the framework of 
the VMD model. 
In the VMD model, the virtual photon in the process $e^-D^+\rightarrow e^-D^+$ 
is coupled to vector mesons $\rho, \omega$, and $J/\psi$, which are
then coupled to the charm meson.
At zero momentum transfer, the following relation holds: 
\begin{eqnarray}
\sum_{V=\rho,\omega,\psi}
\frac{\gamma_V~ g_{V D^+ D^-}}{m_V^2}=e. 
\end{eqnarray}

In the above, $\gamma_V$ is the photon-vector-meson mixing amplitude and
can be determined from the vector meson partial decay width to 
$e^+ e^-$, i.e., 
\begin{eqnarray}
\Gamma_{Vee}=\frac{\alpha\gamma_V^2}{3 m_V^3},   
\end{eqnarray} 
with the fine structure constant $\alpha=e^2/4\pi$. 
The relative signs of $\gamma_V$'s can be determined from 
the hadronic electromagnetic current expressed in terms of quark currents 
\cite{klingl}. 
Since the virtual photon sees the charge of charm quark in the charm meson
through the $\psi DD$ coupling, we have the following relations:
\begin{eqnarray}
\frac{\gamma_\psi g_{\psi D^+ D^-}}{m_\psi^2}=\frac {2}{3}e,~
\frac{\gamma_\rho g_{\rho D^+ D^-}}{m_\rho^2}
+\frac{\gamma_\omega g_{\omega D^+ D^-}}{m_\omega^2}=\frac {1}{3}e. 
\end{eqnarray}
Similarly, one has, from the process $e^-D^0 \rightarrow e^-D^0$, 
\begin{eqnarray}
\frac{\gamma_\psi g_{\psi D^0 \bar {D^0}}}{m_\psi^2}=\frac {2}{3}e,~
\frac{\gamma_\rho g_{\rho D^0 \bar {D^0}}}{m_\rho^2}
+\frac{\gamma_\omega g_{\omega D^0 \bar {D^0}}}{m_\omega^2}=-\frac {2}{3}e. 
\end{eqnarray}
Using $g_{\rho D^+ D^-}=-g_{\rho D^0 \bar {D^0}}=g_{\rho D D}$, 
$g_{\omega D^+ D^-}=g_{\omega D^0 \bar {D^0}}=g_{\omega D D}$, 
and $g_{\psi D^+ D^-}=g_{\psi D^0 \bar {D^0}}=g_{\psi D D}$
from isospin symmetry, we then have
\begin{eqnarray}
\frac{\gamma_\psi g_{\psi D D}}{m_\psi^2}=\frac {2}{3}e,~ 
\frac{\gamma_\rho g_{\rho D D}}{m_\rho^2}
+\frac{\gamma_\omega g_{\omega D D}}{m_\omega^2}=\frac {1}{3}e,~
-&&\frac{\gamma_\rho g_{\rho D D}}{m_\rho^2}
+\frac{\gamma_\omega g_{\omega D D}}{m_\omega^2}=-\frac {2}{3}e.
\label{vmd}
\end{eqnarray} 
From the above equations, we obtain the following coupling constants:
\begin{eqnarray}
g_{\rho D D}=2.52~,~ 
g_{\omega D D}=-2.84~,~
g_{\psi D D}=7.64~.
\end{eqnarray} 
We note that in Ref.\cite{muller} the same VMD relations 
for $g_{\rho D D}$ and $g_{\psi D D}$ as our Eq. (\ref{vmd}) are used 
but slightly different values, i.e., 
$g_{\rho D D}=2.8$ and $g_{\psi D D}=7.7$, are obtained.

Equations similar to Eq. (\ref{vmd}) can be written for kaons and pions 
in order to obtain $g_{VKK}$ and $g_{V\pi \pi}$. 
The resulting coupling constants, multiplied by the corresponding prefactors
in the following SU(4) relations, are given in the 
parentheses for comparison:
\begin{eqnarray}
g_{\rho \pi \pi}(5.04)&& =2 g_{\rho KK}(5.04) 
=2 g_{\rho D D}(5.04) =\frac {\sqrt 6}{2} g_{\psi DD} (9.36)~.
\label{relation}
\end{eqnarray}
We note that $|g_{\rho \pi \pi}|$
is $6.06$ if it is determined from the $\rho$ meson decay width to two pions. 
It is seen that the predicted values differ only slightly from the 
above SU(4) relation except the coupling constant $g_{\psi DD}$. 
This may indicate a sizable uncertainty in the $\psi DD$ coupling. 

\pagebreak
\section*{Appendix B}
\setcounter{equation}{0}
\setcounter{section}{0}
\def\theequation{B\arabic{equation}}

In this appendix, we discuss in detail the differences between our approach 
and that of Ref.\cite{haglin}. 
In particular, 
we compare the general effective Lagrangian for interacting
pseudoscalar and vector mesons and the specific interaction Lagrangians 
for $J/\psi$ scattering by pion and rho meson in the two
approaches. We also examine the condition of current conservation for 
the amplitudes derived from the two approaches.

\subsection{General effective Lagrangian}

In both approaches, one
starts from the same free Lagrangian of Eq. (\ref{lagn0}). 
But our matrices for $P$ and $V$ differ from those of Ref. 
\cite{haglin} by a factor of $1/\sqrt 2$ as given recently in Ref. \cite{pv}.
For the minimal substitutions given in 
Eq. (\ref{ms1}) and Eq. (\ref{ms2}) for obtaining the interaction
Lagrangians, the first one is the same
in the two approaches but the second one is different. Instead of
the factor $g/2$ in the last term of Eq. (\ref{ms2}), Ref. \cite{haglin} 
uses $g$. As a result, the effective Lagrangian given by Eq. (2) in 
Ref.\cite{haglin} has the following form:
\begin{eqnarray}
{\l_{\rm int}}&=& ig {\rm Tr} 
\left ( P V^\mu \partial_\mu P -\partial^\mu P V_\mu P \right ) 
+\frac{1}{2}g^2 {\rm Tr} 
\left ( P V^\mu  V_\mu P- P V^\mu P V_\mu \right ) \nonumber \\
&+& ig {\rm Tr} \left ( \partial^\mu V^\nu \left [V_\mu, V_\nu \right ] 
+\left [V^\mu, V^\nu \right ] \partial_\mu V_\nu \right ) 
+g^2 {\rm Tr} \left ( V^\mu V^\nu \left [V_\mu, V_\nu \right ] \right )~,
\nonumber \\
&=& ig {\rm Tr} \left ( \partial^\mu P \left [P, V_\mu \right ] \right ) 
-\frac{g^2}{4} {\rm Tr}\left (\left [ P, V_\mu \right ]^2 \right )
+ 2ig {\rm Tr} \left ( \partial^\mu V^\nu \left [V_\mu, V_\nu \right ] \right )
+\frac{g^2}{2} {\rm Tr} \left ( \left [V_\mu, V_\nu \right ]^2 \right )~.
\end{eqnarray}
This Lagrangian differs from ours in the three- (VVV) and four- (VVVV)
vector meson couplings. Compared with our Eq. (\ref{lagn2}), 
we find that our VVV and VVVV terms 
are a factor of $2$ and $4$ smaller than corresponding ones
in Ref.\cite{haglin}, respectively.

We note that the above differences in the effective Lagrangians used
in the two approaches cannot be due to a redefinition of fields.
To see this, we rescale the coupling $g$ and the fields $P$ and $V$  
by $c_g, c_P$, and $c_V$, {\em separately},  
then the relative ratios of the PVV, PPVV, VVV, and VVVV terms are given by 
\begin{eqnarray}
c_g c_P^2 c_V,~c_g^2 c_P^2 c_V^2,~2c_g c_V^3,{~\rm and~}4c_g^2 c_V^4.
\end{eqnarray}
It is obvious that these ratios cannot be changed to $1$ simultaneously,
as one needs $c_g c_V=1$ from the ratio of the first two terms  
but $2 c_g c_V=1$ from the ratio of the last two terms. 

\subsection{Lagrangians for the $J/\psi$ interaction with pions and rho mesons}

After expanding the general effective Lagrangian, 
we have Eq. (\ref{pdds}) for the $\pi DD^*$ interaction, which
should be compared with the following one in  
Ref.\cite{haglin}:
\begin{eqnarray}
\l_{\pi DD^*}&=&\frac {i}{2} g^\prime_{\pi DD^*}~ 
\left ( \bar D \tau_i D^{* \mu} \partial_\mu \pi_i - 
\partial_\mu \bar D \tau_i D^*_\mu \pi_i -{\rm H.c.} \right )~,
\label{kh1}
\end{eqnarray}
where we have use $g^\prime$ to label the coupling constant in
Ref.\cite{haglin}. It is seen that
our coupling constant $g_{\pi DD^*}$ is a factor of $2$ smaller.
Apart from the possible difference due to the definition of $D$ field, 
we have the same ${\cal L}_{\pi DD^*}$ as Ref.\cite{haglin}. 
 
Compared to our $\psi DD$ interaction Lagrangian in Eq. (\ref{jdd}), 
Ref.\cite{haglin} has
\begin{eqnarray}
\l_{\psi DD}&=&ig^\prime_{\psi DD}~ \psi^\mu 
\left [ \bar D \partial_\mu D  - (\partial_\mu \bar D) D \right ]. 
\label{kh2}
\end{eqnarray}
Apart from a possible sign difference in the definition of $g_{\psi DD}$,
both have the same ${\cal L}_{\psi DD}$.

Instead of our $\psi D^*D^*$ interaction Lagrangian in Eq. (\ref{jdsds}), 
Ref.\cite{haglin} has
\begin{eqnarray}
\l_{\psi D^*D^*}&=& -ig^\prime_{\psi D^*D^*}~
\psi^\mu \left [ \bar {D^{* \nu}} (\partial_\mu D^*_\nu) 
- (\partial_\mu \bar {D^{* \nu}}) D^*_\nu
- ( \partial_\nu D^*_\mu) \bar {D^{*\nu}}
+ (\partial_\nu \bar {D^*_\mu}) D^{* \nu} \right ].
\label{kh3}
\end{eqnarray}
Besides a possible sign difference in the definition of 
$g_{\psi D^*D^*}$, 
we have two more terms in ${\cal L}_{\psi D^*D^*}$, which involve 
the derivative of the $J/\psi$ field, than in Ref. \cite{haglin}. 
We note that the cyclic form of our ${\cal L}_{\psi D^*D^*}$ 
yields the following factor for the three-vector meson vertex in a Feynman 
diagram:
\begin{eqnarray}
(p_1-p_2)_\gamma g_{\mu \nu}+(p_2-p_3)_\mu g_{\nu \gamma}
+(p_3-p_1)_\nu g_{\gamma\mu},
\end{eqnarray}
which looks exactly like the structure of the three-gluon vertex in QCD. 
A similar difference appears between our approach and that
of Ref. \cite{haglin} for the $\rho D^*D^*$ interaction Lagrangian.
We note that there may be typos in Ref. \cite{haglin} for  
$\l_{\psi D^*D^*}$ and $\l_{\rho D^*D^*}$, as  
$\bar D^* D^*$ and $D^* \bar D^*$ cannot be both scalar.  

For the $\pi \psi DD^*$ interaction Lagrangian, ours is given 
by Eq.(\ref{pjdds}) while that in Ref.\cite{haglin} is
\begin{eqnarray}
\l_{\psi \pi DD^*}&=&-g^\prime_{\psi DD} g^\prime_{\pi DD^*}~
\psi^\mu D^*_\mu \tau_i \bar D \pi_i ~,
\label{kh4}
\end{eqnarray}
which is non-Hermitian and thus likely contains a typo.

\subsection{Current conservation for $\pi\psi\rightarrow D^* \bar D$}

As pointed out in Sec. \ref{sec_cc}, 
in the limit of zero vector meson masses, degenerate pseudoscalar meson
masses, and SU(4) invariant coupling constants, all amplitudes for the
three $J/\psi$ absorption processes shown in Fig. \ref{diagrams} should
satisfy the current conservation condition of Eq. (\ref{current}).
Here, we consider the process $\pi\psi\rightarrow D^* \bar D$ 
as an example and explicitly check the current conservation condition
in both our approach and that of Ref. \cite{haglin}. 
For simplicity, we take all meson masses to be zero. 

\subsubsection{Our approach}

The three amplitudes for the process $\pi \psi \rightarrow D^* \bar D$
are shown in Eq. (\ref{pij}). Multiplying them by $p_{2 \nu}$ 
and omitting the common factor $-g_{\pi D D^*} g_{\psi D D}$, we obtain
\begin{eqnarray}
{\cal M}_{1a}^{\nu \lambda} p_{2 \nu}
&=& \frac {p_2 \cdot (p_1-p_3+p_4)}{t}
(-2p_1+p_3)^\lambda 
= (2p_1-p_3)^\lambda, \nonumber \\
{\cal M}_{1b}^{\nu \lambda} p_{2 \nu} &=& 
\frac {(p_1+p_4)^\alpha}{u} g_{\alpha \beta} 
\left [ (-p_2-p_3)^\beta g^{\nu \lambda}
+(-p_1+p_2+p_4)^\lambda g^{\beta \nu}
+(p_1+p_3-p_4)^\nu g^{\beta \lambda} \right ] p_{2 \nu} \nonumber \\
&=& \frac {(p_1+p_4)_\beta}{u} \left [ -p_3^\beta p_2^\lambda 
+p_2^\beta (-p_1+p_4)^\lambda -u g^{\beta \lambda} \right ] 
=\left [ \left (\frac{-s+t}{2u}\right ) p_3 -p_1-p_4 \right ]^\lambda,
\nonumber \\
{\cal M}_{1c}^{\nu \lambda} p_{2 \nu} &=& p_2^\lambda.
\label{piours}
\end{eqnarray}
Their sum is given by
\begin{eqnarray}
{\cal M}_1^{\nu \lambda} p_{2 \nu} = \left (\frac{-s+t}{2u}\right ) 
~p_3^\lambda \Rightarrow 0.  
\end{eqnarray}
As indicated in the last step, it goes to zero when contracting with 
the external polarization $\epsilon_{3 \lambda}$ of the charm vector meson. 
It is also straightforward to verify the current conservation condition
for $M_1^{\nu \lambda} p_{3\lambda}$ as shown later in Eq. (\ref{p3}).

\subsubsection{Approach of Ref.{\protect \cite{haglin}}}

Using the interaction Lagrangians of Ref. \cite{haglin} as explained
in the above, we obtain the following corresponding amplitudes for
Eqs. (\ref{kh1}), (\ref{kh2}), (\ref{kh3}), and (\ref{kh4}): 
\begin{eqnarray}
m_{1a}^{\nu \lambda}&=& \frac{g^\prime_{\pi D D^*}}{2} g^\prime_{\psi D D}~
(-2p_1+p_3)^\lambda \left (\frac{1}{t-m_D^2} \right ) 
(p_1-p_3+p_4)^\nu, \nonumber \\
m_{1b}^{\nu \lambda}&=& \frac{g^\prime_{\pi D D^*}}{2} g^\prime_{\psi D^* D^*}~
(p_1+p_4)^\alpha \left ( \frac{1}{u-m_{D^*}^2} \right )
\left [ g_{\alpha \beta}-\frac{(p_1-p_4)_\alpha (p_1-p_4)_\beta}{m_{D^*}^2}
\right ] \nonumber \\
&\times& \left [ -p_3^\beta g^{\nu \lambda}
+(-p_1+p_4)^\lambda g^{\beta \nu}
+(p_1+p_3-p_4)^\nu g^{\beta \lambda} \right ] , \nonumber \\
m_{1c}^{\nu \lambda}&=& 
-g^\prime_{\pi D D^*} g^\prime_{\psi D D}~ g^{\nu \lambda}.
\end{eqnarray}
Taking $g^\prime_{\psi D^* D^*}=g^\prime_{\psi D D}$ as in 
Eq. (5) of Ref.\cite{haglin} and omitting the common factor 
$g_{\pi D D^*} g_{\psi D D}/2$, we obtain
\begin{eqnarray}
m_{1a}^{\nu \lambda} p_{2 \nu}
&=& (2p_1-p_3)^\lambda, \nonumber \\
m_{1b}^{\nu \lambda} p_{2 \nu} &=& 
\frac {(p_1+p_4)^\alpha}{u} g_{\alpha \beta} 
\left [ -p_3^\beta g^{\nu \lambda} +(-p_1+p_4)^\lambda g^{\beta \nu}
+(p_1+p_3-p_4)^\nu g^{\beta \lambda} \right ] p_{2 \nu} \nonumber \\
&=& \frac {(p_1+p_4)_\beta}{u} \left [ -p_3^\beta p_2^\lambda 
+p_2^\beta (-p_1+p_4)^\lambda -u g^{\beta \lambda} \right ] 
=\left [ \left (\frac{-s+t}{2u}\right ) p_3 -p_1-p_4 \right ]^\lambda,
\nonumber \\
m_{1c}^{\nu \lambda} p_{2 \nu} &=& -2 p_2^\lambda.
\label{pikh}
\end{eqnarray}
Their sum is thus
\begin{eqnarray}
m_1^{\nu \lambda} p_{2 \nu} = \left [\left (\frac{-s+t}{2u}\right ) p_3
-3p_2 \right ]^\lambda \Rightarrow -3p_2^\lambda. 
\end{eqnarray}
When contracting with the external polarization $\epsilon_{3 \lambda}$
of the charm vector meson, the first term vanishes but the second remains
as shown in the last step. Therefore, the current conservation condition
is not satisfied in Ref. \cite{haglin}.

To understand the above results, we
compare the amplitudes given in Eq. (\ref{pikh}) against
Eq. (\ref{piours}) and note the following two differences. 
(i) In $m_{1b}$ of Eq. (\ref{pikh}), 
the two terms $-p_2^\beta g^{\nu\lambda} +p_2^\lambda g^{\beta\nu}$
involving the four-momentum of $\psi$ in the $\psi D^* D^*$ vertex 
are missing.
(ii) $m_{1c}$ in Eq. (\ref{pikh}) is a factor of $-2$ larger 
than ours [also see the comment after Eq. (\ref{kh4})]. 
In calculating $m_1^{\nu \lambda} p_{2\nu}$, the difference in (i) 
does not matter as it accidentally gives 
$(-p_2^\beta p_2^\lambda +p_2^\lambda p_2^{\beta})=0$.  The failure of 
satisfying the current conservation condition
in Ref.\cite{haglin} is thus due to the difference in (ii).

\begin{figure}
\centerline{\epsfig{file=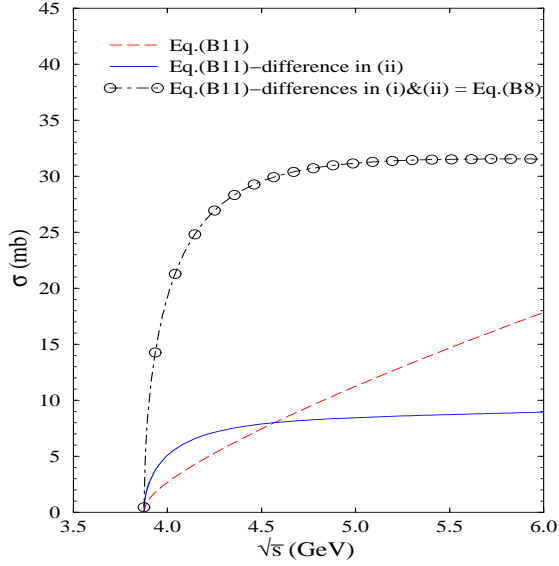,height=3in,width=3in,angle=0}}
\caption{
Comparison of $J/\psi$ absorption cross section by pions. 
The dashed curve represents the result of Eq. (\ref{pikh}) based on
the Lagrangian from Ref.{\protect \cite{haglin}}.
The solid curve is the result from Eq. (\ref{pikh}) by dividing 
the amplitude $m_{1c}$ by a factor of $-2$, which reproduces 
the result from Ref.{\protect \cite{haglin}}.
The circled curve is the result from Eq. (\ref{piours}) based on
our Lagrangian.
}
\label{s-comp}
\end{figure}

To see this more clearly, we show in Fig. \ref{s-comp} by the dashed curve
the cross section for $J/\psi$ absorption by pions 
obtained from the amplitudes given 
in Eq.(\ref{pikh}) and using the coupling constants 
$g^\prime_{\pi D D^*}=8.8$  
and $g^\prime_{\psi D D}=g^\prime_{\psi D^* D^*}=7.7$
given in Ref. \cite{haglin}. However, the results are different from 
that shown in Fig. 1 of Ref. \cite{haglin}, which is reproduced here by
the solid curve in Fig. \ref{s-comp}. We have found that to reproduce
the results in Ref. \cite{haglin} requires dividing the amplitude
$m_{1c}$ in Eq. (\ref{pikh}) by $-2$.

We note that although the current conservation condition 
is satisfied for $m_1^{\nu \lambda} p_{2 \nu}$ when
$m_{1c}$ in Eq. (\ref{pikh}) is divided by a factor of 
$-2$, the current conservation condition for $m_1^{\nu \lambda} 
p_{3 \lambda}$ remains violated. This is shown explicitly in the following.

From the amplitudes shown in Eq. (\ref{pikh}), we have, after 
omitting the common factor $g^\prime_{\pi D D^*} g^\prime_{\psi D D}/2$, 
\begin{eqnarray}
m_{1a}^{\nu \lambda} p_{3 \lambda}
&=& \frac{p_3 \cdot (-2 p_1+p_3)}{t} (p_1-p_3+p_4)^\nu
=(p_1-p_3+p_4)^\nu, \nonumber \\
m_{1b}^{\nu \lambda} p_{3 \lambda} &=& 
\frac {(p_1+p_4)^\alpha}{u} g_{\alpha \beta} 
\left [ -p_3^\beta g^{\nu \lambda} +(-p_1+p_4)^\lambda g^{\beta \nu}
+(p_1+p_3-p_4)^\nu g^{\beta \lambda} \right ] p_{3 \lambda} \nonumber \\
&=& \frac {(p_1+p_4)_\beta}{u} \left [ p_3 \cdot (-p_1+p_4) g^{\nu \beta} 
+p_3^\beta (p_1-p_4)^\nu \right ] 
=\left [\frac{(p_1+p_4)}{-2}
+\left (\frac{s-t}{2u}\right) (p_1-p_4) \right ]^\nu,
\nonumber \\
m_{1c}^{\nu \lambda} p_{3 \lambda} & \rightarrow & 
m_{1c}^{\nu \lambda} p_{3 \lambda}/(-2) =p_3^\nu.
\end{eqnarray}
Their sum is
\begin{eqnarray}
m_1^{\nu \lambda} p_{3 \lambda}=\left [\frac{(p_1+p_4)}{2}
+\left (\frac{s-t}{2u}\right) (p_1-p_4) \right ]^\nu,
\label{b}
\end{eqnarray}
which does not go to zero when contracting with the external polarization 
$\epsilon_{2 \nu}$. 

On the other hand, if we also 
add the two missing terms in $\l_{\psi D^*D^*}$ 
according to our Eq. (\ref{jdsds}), 
i.e., adding $-p_2^\beta g^{\nu\lambda} +p_2^\lambda g^{\beta\nu}$ to 
the $\psi D^*D^*$ vertex in $m_{2b}$ in Eq.(\ref{pikh}), 
we then have the following additional contribution to 
$m_{1b}^{\nu \lambda} p_{3 \lambda}$:
\begin{eqnarray}
\frac {(p_1+p_4)_\beta}{u} \left ( -p_2^\beta p_3^\nu
+p_2 \cdot p_3 g^{\beta ^\nu} \right ) 
=\left [ \frac{(p_1+p_4)}{-2} - \left (\frac{s-t}{2u}\right ) p_3 \right ]^\nu.
\end{eqnarray}
Combining the above two results gives
\begin{eqnarray}
m_1^{\nu \lambda} p_{3 \lambda} =
\left (\frac{s-t}{2u}\right ) (p_1-p_4-p_3)^\nu
=\left (\frac{s-t}{2u}\right ) (-p_2^\nu) \Rightarrow 0.
\label{p3}
\end{eqnarray}
As shown in the last step, it vanishes when contracting with the external 
polarization $\epsilon_{2 \nu}$. The results after eliminating both
differences (i) and (ii), i.e., using our amplitudes
given in Eq. (\ref{piours}), are shown by the circled curve in 
Fig. \ref{s-comp}, which is only slightly different from our 
results shown in Fig. \ref{sigma} due to the different value for 
$g_{\psi D D}$ ($7.7$ vs $7.64$).

\pagebreak
{}


\begin{thebibliography}{99}

\bibitem{matsui}
T. Matsui and H. Satz, \Journal{\PLB}{178}{416}{1986}.
\bibitem{early}
NA38 Collaboration, M. C. Abreu {\it et al.}, 
\Journal{\ZPC}{38}{117}{1988};
C. Baglin {\it et al.}, \Journal{\PLB}{220}{471}{1989}.
\bibitem{su}
C. Baglin {\it et al.}, \Journal{\PLB}{270}{105}{1991};
\Journal{}{345}{617}{1995}.
\bibitem{na50}
NA50 Collaboration, M. Gonin {\it et al.}, 
\Journal{\NP}{A610}{404c}{1996};
NA50 Collaboration, M. C. Abreu {\it et al.}, 
\Journal{\PLB}{450}{456}{1999}.
\bibitem{quant}
D. Kharzeev, C. Louren\c{c}o, M. Nardi, and H. Satz, 
\Journal{\ZPC}{74}{307}{1997}.
\bibitem{review}
See, e.g., R. Vogt, \Journal{\PRT}{310}{197}{1999}.
\bibitem{qgp1}
J.-P. Blaizot and J.-Y. Ollitrault, \Journal{\PRL}{77}{1703}{1996}.
\bibitem{qgp2}
C.-Y. Wong, \Journal{\NP}{A630}{487}{1998}.
\bibitem{conv1}
W. Cassing and C. M. Ko, \Journal{\PLB}{396}{39}{1997};
W. Cassing and E. L. Bratkovskaya, \Journal{\NP}{A623}{570}{1997}.
\bibitem{conv2}
N. Armesto and A. Capella, \Journal{\PLB}{430}{23}{1998}.
\bibitem{cmk}
C. M. Ko, X.-N. Wang, B. Zhang, and X. F. Zhang, 
\Journal{\PLB}{444}{237}{1998}.
\bibitem{redlich}
P. Braun-Munzinger and K. Redlich, 
\Journal{\NP}{A661}{546}{1999}.
\bibitem{blaschke}
K. Martins, D. Blaschke, and E. Quack, \Journal{\PRC}{51}{2723}{1995}.
\bibitem{wong}
C.-Y. Wong, E. S. Swanson, and T. Barnes, hep-ph/9912431.
\bibitem{ope}
D. Kharzeev and H. Satz, \Journal{\PLB}{334}{155}{1994}.
\bibitem{muller}
S. G. Matinyan and B. M\"uller, \Journal{\PRC}{58}{2994}{1998}.
Note that some coupling constants have a factor of 2 
difference in the definitions.
\bibitem{haglin}
K. Haglin, \Journal{\PRC}{61}{031902}{2000}.
\bibitem{ds}
Z. Lin, C. M. Ko, and B. Zhang, 
\Journal{\PRC}{61}{024904}{2000}.
\bibitem{dwidth}
P. Colangelo, F. De Fazio, and G. Nardulli, 
\Journal{\PLB}{334}{175}{1994};
V. M. Belyaev, V. M. Braun, A. Khodjamirian, and R. Ruckl, 
\Journal{\PRD}{51}{6177}{1995}.
\bibitem{form}
J. Vermaseren, computer code {\footnotesize FORM}, 1989.  
The free version of the software 
is available on the internet at 
ftp://hep.itp.tuwien.ac.at/pub/Form/PC/. 
\bibitem{comover}
D. E. Kahana and S. H. Kahana, 
\Journal{\PRC}{59}{1651}{1999};
B.-H. Sa, A. Tai, H. Wang, and F.-H. Liu, 
\Journal{\IBID}{59}{2728}{1999};
C. Spieles, R. Vogt, L. Gerland, S. A. Bass, M. Bleicher, 
H. St\"ocker, and W. Greiner, 
\Journal{\IBID}{60}{054901}{1999}. 
\bibitem{song}
C. Song and V. Koch, \Journal{\PRC}{55}{3026}{1997}.
\bibitem{ff}
J. Kapusta, P. Lichard, and D. Seibert, \Journal{\PRD}{44}{2774}{1991};
\Journal{}{47}{4171(E)}{1993}.
\bibitem{rhoth}
G. E. Brown and M. Rho, \Journal{\PRL}{66}{2720}{1991};
T. Hatsuda and S. H. Lee, \Journal{\PRC}{46}{R34}{1992};
M. Asakawa, C. M. Ko, P. L\'evai, and X. J. Qiu, \Journal{\IBID}{46}{R1159}
{1992}; 
C. M. Ko, V. Koch, and G. Q. Li, Ann. Rev. Nucl. Part. Sci. {\bf 47}, 505
(1997).
\bibitem{rhoexp}
CERES Collaboration, G. Agakishiev {\it et al.}, 
\Journal{\PRL}{75}{1272}{1995};
CERES Collaboration, P. Wurm {\it et al.}, 
\Journal{\NP}{A590}{103c}{1995}.
\bibitem{chan}
L.-H. Chan, \Journal{\PRD}{55}{5362}{1997}.
\bibitem{klingl}
F. Klingl, N. Kaiser, and W. Weise, \Journal{\ZPA}{356}{193}{1996}.
\bibitem{pv}
K. Haglin and C. Gale, nucl-th/0002029.

\end{thebibliography}
\end{document}